# $H_2{}^{16}O$ and $H_2{}^{18}O$ Maser Emission from Gas–Dust Clouds

A. V. Nesterenok* and D. A. Varshalovich

*Ioffe Physical-Technical Institute, ul. Politekhnicheskaya 26, St. Petersburg, 194021 Russia*

**Abstract**: The collisional pumping of $H_2{}^{16}O$ and $H_2{}^{18}O$ masers in hot dense gas–dust clouds has been simulated numerically. New data on the rate coefficients for collisional transitions from Faure et al. (2007) were used in the calculations. The possibility of detecting $H_2{}^{18}O$ emission in 22.2-GHz $H_2{}^{16}O$ maser sources is investigated. The medium is shown to become optically thick in the $H_2{}^{16}O$ lines for which an inverted level population is observed at $H_2O$ column densities of ~$10^{19}$–$10^{20}$ cm$^{-2}$. A simultaneous observation of $H_2{}^{18}O$ emission and $H_2{}^{16}O$ maser emission in the same source will allow the physical conditions in the gas-dust cloud to be refined.

## INTRODUCTION

$H_2{}^{16}O$ emission sources in the $\lambda$ = 1.35 cm (22.23508 GHz) line were discovered by Cheung et al. (1969) in the Orion Nebula and the giant molecular clouds Sgr B2 and W49. A large number of both Galactic and extragalactic maser sources have been detected in this line over the elapsed time. The $H_2{}^{16}O$ maser line at 22.2 GHz corresponds to the permitted electric dipole transition between the rotational levels of ortho-$H_2{}^{16}O$ molecules $J_{KaKc} = 6_{16} \rightarrow 5_{23}$, where $K_a$ and $K_c$ are the asymptotic quantum numbers that characterize the projections of the angular momentum vector **J** onto the molecule's internal axes. Figure 1 presents the energy diagrams of lower rotational levels for the ground electronic-vibrational state of ortho- and para-$H_2{}^{16}O$ molecules. In subsequent theoretical works, level population inversion, along with the $6_{16} \rightarrow 5_{23}$ transition, was predicted for several other $H_2{}^{16}O$ rotational transitions whose frequencies lie in the millimeter and submillimeter wavelength ranges. The emission at these frequencies undergoes strong absorption in the Earth's atmosphere that makes ground-based observations difficult. Using the 10-m telescope of the submillimeter Caltech Observatory on Mauna Kea, Menten et al. (1990) were the first to detect the ortho-$H_2{}^{16}O$ emission at 321 GHz (the $10_{29} \rightarrow 9_{36}$ transition) toward sources in which the 22.2-GHz line emission had been observed previously. The narrowness of the observed lines pointed to maser amplification of this emission. Subsequently, the emission in other "rotational" lines of both ortho-$H_2{}^{16}O$ and para-$H_2{}^{16}O$ of the ground and first excited vibrational states with evidence of maser amplification was discovered (Humphreys 2007).

* E-mail: alex-n10@yandex.ru

As a rule, the collisional excitation of $H_2O$ molecules to higher lying levels followed by the radiative deexcitation of these levels is considered as the main $H_2O$-maser pumping mechanism (Elitzur 1992). In this case, for each photon of the emission being amplified there must be one or more infrared "sink" photons, which must either be absorbed by colder dust or escape from the resonance region to complete the pumping cycle.

The collisional pumping mechanism was considered by Strelnitskii (1973) and de Jong (1973) in the first theoretical $H_2O$-maser models. Apart from the level population inversion in the 22.2-GHz line (the $6_{16} \rightarrow 5_{23}$ transition), de Jong pointed to the possibility of level population inversion for the $5_{32} \rightarrow 4_{41}$ transition of ortho-$H_2{}^{16}O$ at 621 GHz, the $4_{23} \rightarrow 3_{30}$ transition at 448 GHz, and the $4_{14} \rightarrow 3_{21}$ transition at 380 GHz (Fig. 1a).

Deguchi (1981) proposed a collisional 22.2-GHz maser pumping mechanism in which the infrared $H_2O$ line radiation of gas is absorbed by cold dust containing $H_2O$ ice. The absorption coefficient of $H_2O$ ice has a maximum near a wavelength of 43 $\mu$m (230 cm$^{-1}$) and, hence, the $5_{23} \rightarrow 4_{14}$ sink line photons with energy of 221.7 cm$^{-1}$ should be efficiently absorbed.

Varshalovich et al. (1983) showed that the level population inversion is significantly affected by the quasi-resonance excitation energy transfer in collisions of $H_2$ and $H_2O$. Chandra et al. (1984, 1985) investigated the conditions for level population inversion in ortho- and para-$H_2{}^{16}O$ lines. They pointed to the possibility of level population inversion for a number of ortho- and para-$H_2{}^{16}O$ transitions, in particular, for the para-$H_2{}^{16}O$ $3_{13} \rightarrow 2_{20}$ transition at 183 GHz and the $5_{15} \rightarrow 4_{22}$ transition at 325 GHz (Fig. 1b).

The level population inversion for $H_2{}^{16}O$ rotational transitions was studied in detail by Neufeld and Melnick (1991) and Yates et al. (1997). In particular, Yates et al. (1997) pointed to the possibility of level population inversion for 32 ortho-$H_2{}^{16}O$ rotational transitions and 24 para-$H_2{}^{16}O$ transitions of the ground electronic-vibrational state.

Here, we investigate the pumping of $H_2{}^{16}O$ and $H_2{}^{18}O$ masers in dense gas–dust clouds. We consider the collisional $H_2O$ maser pumping mechanism in the presence of cold dust. The rotational level populations are found through a self-consistent solution of the system of balance equations for energy level populations and radiative transfer equations in molecular lines. The goal of this paper is to determine the gains in $H_2{}^{16}O$ and $H_2{}^{18}O$ lines and the physical parameters at which maser amplification of the $H_2{}^{18}O$ emission is possible.

## PHYSICAL PARAMETERS OF THE MEDIUM IN $H_2O$ MASERS

### *The Relative $H_2O$ Abundance in Maser Sources*

According to the data from Lodders et al. (2009), the relative oxygen abundance in the Solar system is O/H = $5.4 \times 10^{-4}$. Adopting this relative oxygen abundance and taking into account the fact that a significant fraction of oxygen can be bound in CO and other chemical compounds, we obtain an estimate for the relative abundance of molecules $H_2O/H_2 \leq 10^{-3}$. In our calculations, we adopted $H_2O/H_2 = 10^{-4}$ (Table 1). Observations of molecular lines in the neighbourhoods of late-type stars and star-forming regions allow the $H_2O$ abundance in these objects to be estimated.

### *The Gas Temperature*

Consider a gas mixture composed of hydrogen and water molecules and their dissociation products. For simplicity, let us assume that there are no other atoms and chemical compounds in the gas mixture. Consider the following chemical reactions:

$$\begin{aligned} H_2 &\leftrightarrow H + H, \\ H_2O &\leftrightarrow H + OH, \\ OH &\leftrightarrow H + O, \\ O_2 &\leftrightarrow O + O. \end{aligned} \tag{1}$$

The degree of dissociation of $H_2$ and $H_2O$ molecules increases with cloud temperature. The degree of molecule dissociation can be determined from a relation similar to Saha's formula for ionization processes:

$$\mathrm{AB} \leftrightarrow \mathrm{A} + \mathrm{B},$$

$$\frac{n_A n_B}{n_{AB}} = \frac{\left(2\pi\mu kT\right)^{3/2}}{h^3} \frac{Z_A Z_B}{Z_{AB}} e^{-D/kT}, \tag{2}$$

where $n_{AB}$, $n_A$, and $n_B$ are the particle number densities, $\mu = m_A m_B/m_{AB}$ is the reduced mass, $D$ is the molecule dissociation energy, $Z_A$, $Z_B$, and $Z_{AB}$ are the partition functions for the atoms and molecules. The $H_2$, $H_2O$, OH, and $O_2$ dissociation energies and the corresponding temperatures are: $D_{H2} = 36\,118.11 \pm 0.08$ cm$^{-1}$, $T_{H2} = D_{H2}/k = 51\,966$ K (Balakrishnan et al. 1992); $D_{H2O} = 41\,145.94 \pm 0.15$ cm$^{-1}$, $T_{H2O} = 59\,200$ K (Maksyutenko et al. 2006); $D_{OH} = 35\,584 \pm 10$ cm$^{-1}$, $T_{OH} = 51\,197$ K (Joens 2001), $D_{O2} = 41\,261 \pm 15$ cm$^{-1}$, $T_{O2} = 59\,365$ K (Brix and Herzberg 1953). The partition function for an atom or molecule is

$$Z = \sum_i g_i \exp\left(-\varepsilon_i/kT\right), \tag{3}$$

where $g_i$ is the degeneracy of energy level $\varepsilon_i$; the summation is over all particle energy levels. The levels with energies $\varepsilon_i >> kT$ make a negligible contribution to $Z$ and the summation in (3) can be limited to a finite number of terms. For the hydrogen and oxygen atoms in these calculations, we adopted $Z_H = 4$ and $Z_O = 9$. The partition function for the hydrogen molecule $Z_{H2}$ was determined from Eq. (3). For the OH, $H_2O$, and $O_2$ molecules, we used the tables of partition functions accessible in the HITRAN 2008 database (Fischer et al. 2003). Let us determine the fractions of molecular hydrogen and oxygen bound in $H_2O$:

$$\xi_H(H_2) = \frac{2n_{H_2}}{N_H},$$

$$\xi_O(H_2O) = \frac{n_{H_2O}}{N_O},$$

where $N_H$ and $N_O$ are the total number densities of hydrogen and oxygen nuclei, respectively:

$$N_H = 2n_{H_2} + n_H + n_{OH} + 2n_{H_2O},$$

$$N_O = n_O + 2n_{O_2} + n_{OH} + n_{H_2O},$$

where $n_H$, $n_O$, $n_{H_2}$, $n_{OH}$, $n_{H_2O}$, and $n_{O_2}$ are the particle number volume densities. For standard cosmic elemental abundances, the number densities $N_H$ and $N_O$ are related by (Lodders et al. 2009)

$$N_O = 5.4\times10^{-4} N_H. \quad (4)$$

Figure 2 presents the results of solving the system of equations including four equations of the form (2) for each of the processes (1) and Eq. (4). The total hydrogen number density $N_H$ is along the horizontal axis. Each solid curve corresponds to a fixed value of the parameter $\xi_O(H_2O)$; each dotted curve corresponds to a fixed value of the parameter $\xi_H(H_2)$. The values of $\xi_O(H_2O)$ and $\xi_H(H_2)$ for each curve are indicated in the figure. At hydrogen number densities in the cloud $N_H > 10^8$ cm$^{-3}$, the dissociation of water becomes significant only at gas temperatures above 1500 K. The derived values of $\xi_O(H_2O)$ should be considered as an upper limit.

*The Sizes of Maser Sources*

According to the radio-interferometric observations of maser sources in the envelopes of late-type stars and in star-forming regions, the sizes of individual maser clumps are $10^{13}$–$10^{14}$ cm (Bains et al. 2003; Matveyenko et al. 2005). To all appearances, the sizes of maser clumps in the accretion disks in the central regions of active galactic nuclei are of the same order of magnitude (Kartje et al. 1999).

*The Beaming Angle of Maser Emission*

The beaming solid angle of maser emission $\Delta\Omega$ is determined by the geometry and the gas velocity field in the emitting cloud. Based on numerical simulations and analysis of the spectral intensity profile for the maser emission from Galactic sources, Anderson and Watson (1993) concluded that $\Delta\Omega \leq 10^{-5}$–$10^{-4}$ sr.

The maser emission from the central region of NGC 4258 originates in a thin, differentially rotating accretion disk seen edge-on (Moran 2008). The disk thickness in the part where the maser emission is observed is $h \sim 10^{15}$ cm (Argon et al. 2007). The gain length, the distance at which the radiation either escapes from the disk plane or escapes from resonance due to differential disk rotation, is $l \sim 10^{17}$ cm (Nesterenok and Varshalovich 2010). Hence we can obtain an estimate for the beaming angle of the resonance maser emission from the central source, $\Delta\Omega \sim h^2/l^2 \sim 10^{-4}$ sr.

*The Dust Model*

The transition frequencies between the rotational $H_2O$ levels of the ground vibrational state usually do not exceed $5\times10^4$ GHz (6 $\mu$m) and correspond to the infrared and radio bands. The dependences of the dust absorption coefficient on radiation wavelength in the long-wavelength infrared range can be fitted by a power law (Mathis 1990):

$$\kappa_\nu^{(c)} \propto \left(\frac{\nu}{\nu_0}\right)^p,$$

where $\nu$ is the radiation frequency, the superscript ($c$) denotes the absorption of radiation in continuum. The exponent $p$ in different models varies between 0 and 2. In our calculations, we used the dust model from Mathis (1990):

$$\kappa_\nu^{(c)} = 2.1\times10^{-25}\,\text{cm}^2 \times \left(\frac{250\,\mu\text{m}}{\lambda}\right)^2 N_H, \qquad \lambda > 250\,\mu\text{m},$$

where $N_{\text{H}}$ is the total number density of hydrogen nuclei in $\text{cm}^{-3}$ and $\lambda = c/\nu$ is the radiation wavelength. For wavelengths 20 $\mu$m $< \lambda <$ 250 $\mu$m, the dust absorption coefficient was fitted by a piecewise linear function. For wavelengths $\lambda <$ 20 $\mu$m, the absorption coefficient was taken to be

$$\kappa_\nu^{(c)} = 1.0\times10^{-23}\,\text{cm}^2 \times N_H, \quad \lambda < 20\,\mu\text{m}.$$

The dust emissivity was defined by the expression

$$\varepsilon_\nu^{(c)} = \kappa_\nu^{(c)} \times \frac{2h\nu}{\lambda^2}\frac{1}{\exp\left(h\nu/kT_d\right)-1},$$

where $T_d$ is the dust temperature. In our calculations, we assumed the dust temperature to be 100 K; at $T_d$ much lower than the gas kinetic temperature, the dust radiation has virtually no effect on the $H_2O$ rotational level populations.

Table 1 gives the values of the physical parameters used in our numerical calculations.

## THE SCHEME OF CALCULATIONS

*The Cloud Model*

If the optical density of dust in a cloud is much smaller than unity, then the escape of radiation from the resonance region can be determined by both finite cloud sizes and the existence of a gas velocity gradient in the cloud. In the latter case, the radiation escapes from resonance when the frequency difference between the radiation field and intrinsic gas radiation exceeds the molecular line width. Consider a cloud model in which the length of the resonance region or the cloud sizes along two coordinate axes is much larger than those along the third axis (Fig. 3). Such a situation can take place in the accretion disks around compact massive objects and in the gas clouds formed at shock fronts.

*Radiative Transfer in the Cloud*

Let us write the radiative transfer equation

$$\frac{dI_\nu(\boldsymbol{n})}{dl} = -\kappa_\nu I_\nu(\boldsymbol{n}) + \varepsilon_\nu \qquad (5)$$

$$I_\nu(\boldsymbol{n}, l=0) = I_{0\nu}(\boldsymbol{n}),$$

where $I_\nu(\mathbf{n})$ is the intensity of radiation at frequency $\nu$ in a given direction $\mathbf{n}$, $dl$ is the differential of the path length in the direction $\mathbf{n}$, $\varepsilon_\nu$ is the emission coefficient, and $\kappa_\nu$ is the absorption coefficient. Each of the coefficients $\varepsilon_\nu$ and $\kappa_\nu$ is the sum of the emission or absorption coefficients in continuum and in a spectral line, respectively,

$$\varepsilon_\nu = \varepsilon_\nu^{(c)} + \frac{h\nu}{4\pi} A_{ik} \phi_\nu n_i N,$$

$$\kappa_\nu = \kappa_\nu^{(c)} + \frac{\lambda^2}{8\pi} A_{ik} \left( \frac{g_i}{g_k} n_k - n_i \right) N \phi_\nu,$$

where $\varepsilon_\nu^{(c)}$ and $\kappa_\nu^{(c)}$ are the emission and absorption coefficients in continuum, $A_{ik}$ is the Einstein coefficient for spontaneous radiation, $n_i$ and $n_k$ are the normalized populations of levels $i$ and $k$, $\sum_j n_j = 1$, $N$ is the particle number density, $g_i$ and $g_k$ are the statistical weights of the levels, $\nu$ and $\lambda$ are the radiation frequency and wavelength, and $\varphi_\nu$ is the normalized spectral profile. In these formulas, it is implied that level $i$ lies above level $k$ in energy, i.e., $\varepsilon_i > \varepsilon_k$. The spectral profile of the emission and absorption coefficients $\varphi_\nu$ is

$$\phi_\nu = \frac{1}{\sqrt{\pi}\Delta\nu} \exp\left( -\frac{(\nu - \nu_{ik})^2}{(\Delta\nu)^2} \right),$$

where $\Delta\nu$ is the profile width and $\nu_{ik}$ is the mean frequency of the transition from level $i$ to level $k$. The relation between the Einstein coefficient $A_{ik}$ and the matrix element of the dipole moment operator for the transition is

$$g_i A_{ik} = \frac{4}{3}(2\pi/\lambda)^3 \frac{2\pi S_{ik}}{h}, \qquad (6)$$

where $S_{ik} = \left|\left\langle k \middle| \hat{d} \middle| i \right\rangle\right|^2$ is the square of the reduced transition matrix element of the dipole moment operator, $h$ is the Planck constant.

Let us choose a coordinate system in such a way that the $z$ axis is parallel to the normal to the cloud plane (Fig. 3). Instead of the geometrical depth along the $z$ axis, we will introduce the optical depth

$$\tau_\nu^{(z)} = \int_0^z \kappa_\nu(z')\,dz'.$$

Since the geometry is symmetric, the solution of Eq. (5) for the intensity does not depend on the azimuth angle. The intensity of radiation at depth $\tau_\nu^{(z)}$ for a direction characterized by an angle $\theta$ to the $z$ axis is

$$I_\nu(\theta) = I_{0\nu} \exp\left(-\frac{\tau_\nu^{(z)}}{\cos\theta}\right) + \int_0^{\tau_\nu^{(z)}} S_\nu\left(\tau_\nu'^{(z)}\right) \exp\left(-\frac{\tau_\nu^{(z)} - \tau_\nu'^{(z)}}{\cos\theta}\right) \frac{d\tau_\nu'^{(z)}}{\cos\theta}, \quad \theta < \frac{\pi}{2},$$

$$I_\nu(\theta) = I_{0\nu} \exp\left(-\frac{\tau_{0\nu}^{(z)} - \tau_\nu^{(z)}}{|\cos\theta|}\right) + \int_{\tau_\nu^{(z)}}^{\tau_{0\nu}^{(z)}} S_\nu\left(\tau_\nu'^{(z)}\right) \exp\left(-\frac{\tau_\nu'^{(z)} - \tau_\nu^{(z)}}{|\cos\theta|}\right) \frac{d\tau_\nu'^{(z)}}{|\cos\theta|}, \quad \theta > \frac{\pi}{2}, \qquad (7)$$

where $\tau_{0\nu}^{(z)}$ is the total optical depth of the cloud at frequency $\nu$ along the $z$ axis (Fig. 3), $S_\nu$ is the source function, $S_\nu = \varepsilon_\nu/\kappa_\nu$, and $I_{0\nu}$ is the intensity of the background radiation. Formulas (7) also hold for a negative absorption coefficient; in this case, $S_\nu$ and $\tau_\nu$ are also negative. Below, we will assume that the background radiation is much less intense than the intrinsic gas radiation and the first term in (7) may be neglected.

*Calculating the Populations of $H_2O$ Rotational Levels*

In the stationary case, the system of equations for the level populations is

$$\sum_{\substack{k=1 \\ k\neq i}}^{M} n_k (R_{ki} + C_{ki}) - n_i \sum_{\substack{k=1 \\ k\neq i}}^{M} (R_{ik} + C_{ik}) = 0, \qquad \sum_{i=1}^{M} n_i = 1, \qquad (8)$$

where $n_i$ and $n_k$ are the populations of levels $i$ and $k$, $M$ is the total number of levels, $R_{ik}$ are the rate coefficients for radiative transitions from level $i$ to level $k$, and $C_{ik}$ are the rate coefficients for collisional transitions. The rate coefficients for radiative transitions $R_{ik}$ are

$$R_{ik}^{\downarrow} = B_{ik}\bar{J}_{ik} + A_{ik}, \ \varepsilon_i > \varepsilon_k,$$

$$R_{ik}^{\uparrow} = B_{ik}\bar{J}_{ik}, \ \varepsilon_i < \varepsilon_k.$$

where $\varepsilon_i$ and $\varepsilon_k$ are the level energies, $A_{ik}$ and $B_{ik}$ are the Einstein coefficients for spontaneous and stimulated radiation, $\bar{J}_{ik}$ is the radiation intensity averaged over the direction and over the line profile:

$$\bar{J}_{ik} = \int_0^{4\pi} \frac{d\Omega}{4\pi} \int_0^{\infty} d\nu\, \phi_\nu I_\nu(\boldsymbol{n}),$$

$I_\nu(\mathbf{n})$ is the solution of Eq. (5). The Einstein coefficients for spontaneous and stimulated radiation $A_{ik}$, $B_{ik}$, and $B_{ki}$ are related by the well-known relations

$$A_{ik} \frac{\lambda^2}{2h\nu} = B_{ik} = \frac{g_k}{g_i} B_{ki},\ \varepsilon_i > \varepsilon_k .$$

The following relation holds for the rate coefficients of collisional excitation and deexcitation:

$$C_{ik} = \frac{g_k}{g_i} C_{ki} \exp\left(\frac{\varepsilon_i - \varepsilon_k}{kT}\right),$$

where $T$ is the gas kinetic temperature.

We will assume that the physical parameters of the cloud (the temperature, gas and dust number densities) are the same for the entire cloud. We will also assume the distribution of molecules in energy levels to be the same for the entire cloud. In this case, the source function is a constant and can be taken outside the integral sign in (7). For a positive absorption coefficient, the intensity averaged over the direction and over the line profile in the central cloud plane is

$$\bar{J}_{ik} = \int d\nu\, \phi_\nu S_\nu \left(1 - \int_0^1 d\mu \exp\left(-\frac{\tau_\nu}{\mu}\right)\right), \qquad (9)$$

$$\tau_\nu = \tau_{0\nu}/2 = \kappa_\nu d/2,$$

where $d$ is the geometrical thickness of the cloud. Equation (9) cannot be used for a negative absorption coefficient, because, in this case, the finite cloud sizes should be taken into account. In our simulations of the pumping mechanism we disregarded the radiation of inverted transitions.

The system of equations (8) is a system of nonlinear equations for the level populations $n_i$, $i = 1, 2, ..., M$. This system was solved by the Newton–Raphson method (Press et al. 1992). The Boltzmann distribution or the solution for an optically thin medium was used as the initial approximation of the solution for the level populations. The integrals in Eq. (9) were calculated by numerical methods.

## SPECTROSCOPIC DATA AND RATE COEFFICIENTS FOR $H_2{}^{16}O$ AND $H_2{}^{18}O$ COLLISIONAL TRANSITIONS

We calculated the level populations for the lowest 45 rotational levels of the ground electronic-vibrational state of ortho-$H_2O$ and the lowest 45 rotational levels of the ground electronic-vibrational state of para-$H_2O$. The energy of the uppermost level is about 1400 $cm^{-1}$, or 2000 K in units of temperature.

The spectroscopic data for the $H_2{}^{16}O$ and $H_2{}^{18}O$ molecules were taken from the HITRAN 2008 database (Rothman et al. 2009). The rate coefficients for $H_2O$ collisional transitions in inelastic collisions of $H_2O$ with He were taken from Green et al. (1993). The helium-to-hydrogen ratio was assumed to be He/H = 0.07. The rate coefficients for $H_2O$ collisional transitions in collisions of $H_2O$ with $H_2$ were taken from Faure et al. (2007). The rate coefficients in Faure et al. (2007) were calculated by the method of quasi-classical trajectories. Note that the difference between the rate coefficients obtained by this method and the results of quantum-mechanical calculations can reach in some cases a factor of 3 (Dubernet et al. 2009). The ortho-to-para-$H_2$ ratio in our calculations was assumed to be equal to three.

The rate coefficients for $H_2O$ collisional transitions in inelastic collisions of $H_2O$ with $H_2$ from Faure et al. (2007) and Dubernet et al. (2009) exceed those obtained from the rate coefficients by Green et al. (1993). This imposes new constraints on the possible molecular hydrogen number densities in maser clouds. The level populations are thermalized at lower $H_2$ number densities than was thought previously (Yates et al. 1997).

## RESULTS OF OUR CALCULATIONS OF THE LEVEL POPULATION INVERSION IN AN UNSATURATED MASER

### *Results of Our Calculations for $H_2{}^{16}O$*

In the presence of an inverted level population, the absorption coefficient in a line becomes negative, $\kappa_\nu < 0$. Below, we will consider only the transitions for which $\kappa_\nu^{(c)} << |\kappa_\nu|$. Let us define the gain $\gamma_\nu = -\kappa_\nu$ ; the expression for $\gamma_\nu$ in an unsaturated maser is

$$\gamma_\nu = \frac{\lambda^2}{8\pi} A_{ik} \left( n_i - \frac{g_i}{g_k} n_k \right) N \times \frac{1}{\sqrt{\pi}\Delta\nu} \exp\left( -\frac{(\nu - \nu_{ik})^2}{(\Delta\nu)^2} \right),$$

$$\Delta\nu = \nu_{ik} \frac{\upsilon}{c},$$

where $n_i$ and $n_k$ are the populations of the upper and lower levels normalized to unity, $N$ is the number density of “working” molecules, $\upsilon$ is a quantity characterizing the velocity spread due to the thermal motion of molecules and the turbulent motions of gas in the cloud,

$$\upsilon^2 = \upsilon_T^2 + \upsilon_{turb}^2,$$

$$\upsilon_T = \sqrt{\frac{2kT}{m_{H_2O}}} = \alpha \sqrt{\frac{T}{1000\,K}} \ \ \mathrm{km\,s^{-1}},$$

where $\alpha = 0.96$ for $H_2{}^{16}O$ and $\alpha = 0.91$ for $H_2{}^{18}O$.

Since the total nuclear spin of ortho-$H_2O$ is equal to unity, the molecular levels are split into three sublevels, $F = J - 1, J, J + 1$, where $F$ and $J$ are the quantum numbers that characterize the total and rotational angular momenta of the molecule, respectively. For the 22.2-GHz ortho-$H_2{}^{16}O$ line, the spectral profile of the emission and absorption coefficients will be the sum of six components with different intensities (Varshalovich et al. 2006). The components $F' \to F = 7 \to 6$, $6 \to 5$, and $5 \to 4$ are most intense; their relative intensities are 0.3919, 0.3302, 0.2779, respectively.

Figure 4 presents the spectral gain profile for $T = 50$, 150, 400, and 1000 K. The frequency shift of the components $7 \to 6$ and $5 \to 4$ in units of relative velocity is 1.029 km s$^{-1}$ and is comparable to the Doppler width of the spectral profile. At $T > 150$ K, the three intense components merge into one asymmetric profile and become unresolvable. When the full width of the spectral gain profile in the 22.2-GHz line is calculated, the additional profile broadening due to the hyperfine splitting should be taken into account:

$$\upsilon^2 \approx \left(\upsilon_T + 0.5\,\mathrm{km\,s^{-1}}\right)^2 + \upsilon_{turb}^2 .$$

For the remaining ortho-$H_2{}^{16}O$ transitions, the splitting is small compared to the Doppler profile width.

Tables 2a and 3a list the parameters of the ortho- and para-$H_2{}^{16}O$ transitions for which, according to our calculations, level population inversion is observed. Columns 2, 3, 4, and 5 give the frequencies $v$, wavelengths $\lambda$, Einstein coefficients $A_{ik}$, and $2\pi S_{ik}/h$ for the transitions under consideration. Tables 2b and 3b list the gains at the line center and per unit number density of “working” molecules:

$$\chi_{ik} = \frac{\lambda^3}{8\pi\sqrt{\pi}\upsilon} A_{ik}\left(n_i - \frac{g_i}{g_k} n_k\right), \text{ or}$$

$$\chi_{ik} = \frac{4}{3}\pi^{3/2}\frac{2\pi S_{ik}}{\upsilon h}\left(\frac{n_i}{g_i} - \frac{n_k}{g_k}\right). \quad (10)$$

In deriving the second expression for $\chi_{ik}$, we used Eq. (6). According to (10), the parameter $\chi_{ik}$ for a given transition is determined by the magnitude of the square of the reduced transition matrix element of the dipole moment operator $S_{ik}$ and the level population inversion ($n_i/g_i - n_k/g_k$). The gain at the line center is $\chi_{ik}N$.

*Results of Our Calculations for $H_2{}^{18}O$*

According to Ball et al. (1971), the frequencies of the three most intense hyperfine components of the ortho-$H_2{}^{18}O$ $6_{16} \to 5_{23}$ line are 5625.105 ± 0.005 MHz for $F' \to F = 7 \to 6$, 5625.139 ± 0.005 MHz for $6 \to 5$, and 5625.183 ± 0.005 MHz for $5 \to 4$. The relative intensities of the components are the same as those for the ortho-$H_2{}^{16}O$ $6_{16} \to 5_{23}$ transition. However, the relative splitting of the components expressed in units of velocity is more than 1.5 km $s^{-1}$, which exceeds the Doppler line broadening of about 1 km $s^{-1}$. In this case, the spectral gain profile consists of the resolved profiles of each of the splitting components. For this line, we considered the radiation of only the most intense hyperfine component in our calculations.

Tables 4a and 4b present the results of our numerical calculations of the level population inversion for $H_2{}^{18}O$ transitions. The $H_2{}^{18}O$-to-$H_2{}^{16}O$ abundance ratio was assumed to be 0.002, the mean value in the Solar system (Lodders et al. 2009); the ortho-to-para-$H_2{}^{18}O$ ratio was assumed to be equal to three. Only the transitions for which $\chi_{ik}$ exceeds $10^{-18}$ $cm^2$ are listed. Six columns in both tables give the $H_2{}^{16}O$ column densities at which the optical depth in $H_2{}^{18}O$ lines is equal to unity.

## THE LIMITING "PHOTON" LUMINOSITY OF A SATURATED MASER

Let us define the photon luminosity of a maser $\Phi_{ik}$ as the number of photons in the maser line emitted by 1 $cm^3$ of gas per 1 s. As the degree of maser saturation increases, the photon luminosity approaches its limiting value. According to Neufeld and Melnick (1991), the expression for the limiting photon luminosity of a saturated maser is

$$\Phi_{ik,\mathrm{lim}} = \frac{n_i/g_i - n_k/g_k}{(\Gamma_i g_i)^{-1} + (\Gamma_k g_k)^{-1}} N,$$

where $n_i$ and $n_k$ are the populations of the upper and lower levels in an unsaturated maser, $N$ is the number density of "working" molecules, $\Gamma_i$ and $\Gamma_k$ are the signal level decay rates:

$$\Gamma_i = \sum_{l \neq k,i} R_{il} + \sum_{l \neq i} C_{il},$$

where the summation includes all of the collisional and radiative transitions to other levels, except the radiative transitions in the maser line. Let us define $f_{ik} \equiv \Phi_{ik,\mathrm{lim}}/N$, the limiting number of photons in the maser line emitted by one molecule per 1 s. The values of the parameter $f_{ik}$ for the transitions under consideration are given in Tables 2b, 3b, 4a, and 4b.

## DISCUSSION

According to our numerical calculations, the gain in the 22.2-GHz ortho-$H_2{}^{16}O$ line for a wide range of physical parameters turns out to be lower than the gains in the millimeter and submillimeter $H_2{}^{16}O$ lines for unsaturated masers (Tables 2b and 3b). This is partly due to the fact that the spectral gain profile in the 22.2-GHz line is additionally broadened by the hyperfine level splitting. In addition, the matrix element of the dipole moment operator for the $6_{16} \rightarrow 5_{23}$ transition has the lowest value among the transitions under consideration (Tables 2a and 3a).

A maser reaches saturation when the rate coefficients for induced radiative transitions between signal levels become comparable to those for radiative and collisional transitions from the signal levels to other energy levels. According to our estimates, the optical depth at which a 22.2-GHz maser becomes saturated is about 30–35, while for masers in the millimeter and submillimeter wavelength ranges it is about 20–25. The limiting photon luminosity of a maser in the 22.2-GHz line is one of the highest for the transitions under consideration (Tables 2b and 3b). This explains the high luminosity of emission sources in the 22.2-GHz line.

Since the number density of $H_2^{18}O$ molecules is lower, the optical depth in "rotational" $H_2^{18}O$ lines is smaller than that for $H_2^{16}O$. The upper $H_2^{18}O$ energy levels have lower populations than the corresponding $H_2^{16}O$ levels. As a result, the collisional pumping of $H_2^{18}O$ masers is less efficient, which is reflected in lower values of the parameter $f_{ik}$ for most transitions (Tables 4a and 4b).

The gains per unit number density of "working" molecules $\chi_{ik}$ for some $H_2^{18}O$ transitions are higher than those for $H_2^{16}O$. Nevertheless, high $H_2O$ column densities, $\sim 10^{19}$–$10^{20}$ cm$^{-2}$ (Tables 4a and 4b), are needed for the optical depth in the $H_2^{18}O$ lines for which level population inversion is observed to exceed 1. The detection of $H_2^{18}O$ maser emission in $H_2^{16}O$ maser sources would be indicative of high $H_2O$ column densities.

Ball et al. (1971) undertook a search for $H_2^{18}O$ maser emission in the 5.63-GHz line (the $6_{16} \rightarrow 5_{23}$ transition) at the 37-m Haystack radio telescope. No 5.63-GHz emission was detected in the then known Galactic $H_2^{16}O$ maser sources. According to our numerical calculations, the optical depth in the 5.63-GHz line reaches unity at $H_2O$ column densities of $\sim 10^{20}$ cm$^{-2}$ (Table 4a).

The existing theoretical models of $H_2O$ masers predict $H_2O$ column densities along the line of sight as high as $10^{20}$ cm$^{-2}$ (Chandra et al. 1985; Yates et al. 1997). According to these calculations, the optical depth at such column densities can reach 10 in the lines of ortho-$H_2^{18}O$ at 390 GHz ($4_{14} \rightarrow 3_{21}$), 489 GHz ($4_{23} \rightarrow 3_{30}$), and 1181 GHz ($3_{12} \rightarrow 2_{21}$), and para-$H_2^{18}O$ at 203 GHz ($3_{13} \rightarrow 2_{20}$). Investigating the $H_2^{18}O$ emission parameters with a high angular resolution in $H_2^{16}O$ maser sources would allow constraints to be imposed on the physical parameters of maser clumps. These data will help improve the theoretical models for the $H_2O$ maser pumping and polarization mechanisms.

## CONCLUSIONS

We numerically simulated the pumping of $H_2^{16}O$ and $H_2^{18}O$ masers. The spectral gain profile in the 22.2-GHz ortho-$H_2^{16}O$ line was additionally broadened by the hyperfine level splitting. Allowance for this splitting leads to a decrease in the effective gain by a factor of 1.2–1.5. For the remaining ortho-$H_2^{16}O$ transitions, the hyperfine splitting is small compared to the Doppler profile width. According to our numerical calculations, the gain in the 22.2-GHz line is lower than the gains in the millimeter and submillimeter $H_2{}^{16}O$ lines in unsaturated masers. We also considered the possibility of detecting $H_2{}^{18}O$ emission in $H_2{}^{16}O$ maser sources.

ACKNOWLEDGMENTS

This work was supported by the Russian Foundation for Basic Research (project no. 11-02-01018), the Program of the President of Russia for Support of Leading Scientific Schools (project no. NSh-3769.2010.2), the Program of the Division of Physical Sciences of the Russian Academy of Sciences (project OFN-16), and the Ministry of Education and Science of the Russian Federation (contract no. 11.G34.31.0001). We are grateful to L.I. Matveyenko for helpful remarks.

**Table 1.** Physical parameters adopted in our calculations

| Parameter | Value |
|---|---|
| Gas kinetic temperature | 400 K, 1000 K |
| Molecular hydrogen number density | $10^8$ cm$^{-3}$ |
| Relative abundance of molecules $H_2O/H_2$ | $10^{-4}$ |
| Ortho-to-para-$H_2O$ ratio | 3 |
| Cloud thickness along normal to cloud plane | $10^{14}$ cm |
| Velocity of turbulent motions in cloud | 1 km s$^{-1}$ |
| Dust temperature | 100 K |

**Table 2a**. Ortho-$H_2{}^{16}O$ transitions for which an inverted level population may be observed according to our calculations

| No. | Transition $i \to k$ | Frequency, GHz | Wavelength, mm | $A_{ik}$, $s^{-1}$ | $2\pi S_{ik}/h$, $10^{-10}$ $s^{-1}$ $cm^3$ |
|---|---|---|---|---|---|
| | 1 | 2 | 3 | 4 | 5 |
| 1 | $6_{16} \to 5_{23}$ | 22.235080 | 13.5 | $1.99\times10^{-9}$ | 1.92 |
| 2 | $10_{29} \to 9_{36}$ | 321.22555 | 0.933 | $6.12\times10^{-6}$ | 3.16 |
| 3 | $4_{14} \to 3_{21}$ | 380.19748 | 0.789 | $3.06\times10^{-5}$ | 4.09 |
| 4 | $6_{43} \to 5_{50}$ | 439.15098 | 0.683 | $2.78\times10^{-5}$ | 3.48 |
| 5 | $4_{23} \to 3_{30}$ | 448.00119 | 0.669 | $5.41\times10^{-5}$ | 4.41 |
| 6 | $5_{32} \to 4_{41}$ | 620.70104 | 0.483 | $1.09\times10^{-4}$ | 4.10 |
| 7 | $6_{34} \to 5_{41}$ | 1158.3240 | 0.259 | $1.40\times10^{-3}$ | 9.53 |
| 8 | $8_{54} \to 7_{61}$ | 1168.3585 | 0.257 | $9.31\times10^{-4}$ | 8.08 |
| 9 | $7_{43} \to 6_{52}$ | 1278.2660 | 0.235 | $1.53\times10^{-3}$ | 8.94 |
| 10 | $8_{27} \to 7_{34}$ | 1296.4113 | 0.231 | $1.06\times10^{-3}$ | 6.71 |
| 11 | $6_{25} \to 5_{32}$ | 1322.0651 | 0.227 | $2.30\times10^{-3}$ | 10.6 |
| 12 | $8_{45} \to 7_{52}$ | 1884.8880 | 0.159 | $7.26\times10^{-3}$ | 15.0 |
| 13 | $8_{36} \to 7_{43}$ | 2244.8111 | 0.134 | $1.47\times10^{-2}$ | 18.0 |

Note. For the $6_{16} \to 5_{23}$ transition, the table gives the mean frequency $\nu_{ik}$ = 22.235080 GHz; the frequencies of the three most intense hyperfine splitting components are $\nu$ = 22.235044 GHz for $F' \to F = 7 \to 6$, 22.235077 GHz for $F' \to F = 6 \to 5$, and 22.235120 GHz for $F' \to F = 5 \to 4$ (Varshalovich et al. 2006).

**Table 2b**. Parameters $\chi_{ik}$ and $f_{ik}$ for ortho-$H_2{}^{16}O$ lines.

| No. | Transition $i \to k$ | 400 K | | 1000 K | |
|---|---|---|---|---|---|
| | | $\chi_{ik}$, $10^{-18}$ $cm^2$ | $f_{ik}$, $10^{-2}$ $s^{-1}$ | $\chi_{ik}$, $10^{-18}$ $cm^2$ | $f_{ik}$, $10^{-2}$ $s^{-1}$ |
| | 1 | 2 | 3 | 4 | 5 |
| 1 | $6_{16} \to 5_{23}$ | 2.8 | 0.6 | 4.4 | 2.4 |
| 2 | $10_{29} \to 9_{36}$ | 1.2 | 0.3 | 5.7 | 2.4 |
| 3 | $4_{14} \to 3_{21}$ | 5.5 | 0.2 | 13 | 1.5 |
| 4 | $6_{43} \to 5_{50}$ | 1.0 | 0.3 | 3.1 | 1.9 |
| 5 | $4_{23} \to 3_{30}$ | 2.2 | 0.15 | 9.6 | 1.6 |
| 6 | $5_{32} \to 4_{41}$ | 2.6 | 0.4 | 8.7 | 2.6 |
| 7 | $6_{34} \to 5_{41}$ | - | - | 13 | 2.3 |
| 8 | $8_{54} \to 7_{61}$ | - | - | 0.6 | 0.2 |
| 9 | $7_{43} \to 6_{52}$ | 0.3 | 0.03 | 6.5 | 1.6 |
| 10 | $8_{27} \to 7_{34}$ | 3.8 | 0.3 | 12 | 2.6 |
| 11 | $6_{25} \to 5_{32}$ | - | - | 14 | 1.5 |
| 12 | $8_{45} \to 7_{52}$ | - | - | 0.9 | 0.1 |
| 13 | $8_{36} \to 7_{43}$ | - | - | 3.4 | 0.4 |

Note. To obtain the gain $\gamma_\nu$ at the line center, the parameter $\chi_{ik}$ should be multiplied by the number density of "working" molecules - ortho-$H_2{}^{16}O$. The dash implies the absence of an inverted level population. The parameters used in our calculations are listed in Table 1.

**Table 3a**. Para-$H_2{}^{16}O$ transitions for which an inverted level population may be observed according to our calculations

| № | Transition $i \to k$ | Frequency, GHz | Wavelength, mm | $A_{ik}$, $s^{-1}$ | $2\pi S_{ik}/h$, $10^{-10}$ $s^{-1}$ $cm^3$ |
|---|---|---|---|---|---|
| | 1 | 2 | 3 | 4 | 5 |
| 1 | $3_{13} \to 2_{20}$ | 183.3101 | 1.64 | $3.63\times10^{-6}$ | 3.36 |
| 2 | $5_{15} \to 4_{22}$ | 325.15310 | 0.922 | $1.16\times10^{-5}$ | 3.03 |
| 3 | $6_{42} \to 5_{51}$ | 470.88908 | 0.637 | $3.44\times10^{-5}$ | 3.49 |
| 4 | $5_{33} \to 4_{40}$ | 474.68928 | 0.632 | $4.76\times10^{-5}$ | 3.99 |
| 5 | $9_{28} \to 8_{35}$ | 906.20596 | 0.331 | $2.18\times10^{-4}$ | 4.54 |
| 6 | $4_{22} \to 3_{31}$ | 916.17142 | 0.327 | $5.69\times10^{-4}$ | 5.42 |
| 7 | $5_{24} \to 4_{31}$ | 970.31525 | 0.309 | $8.97\times10^{-4}$ | 8.80 |
| 8 | $7_{44} \to 6_{51}$ | 1172.5260 | 0.256 | $1.17\times10^{-3}$ | 8.83 |
| 9 | $7_{26} \to 6_{33}$ | 1440.7820 | 0.208 | $2.27\times10^{-3}$ | 9.27 |
| 10 | $6_{33} \to 5_{42}$ | 1541.9671 | 0.194 | $3.60\times10^{-3}$ | 10.4 |
| 11 | $7_{35} \to 6_{42}$ | 1766.1989 | 0.170 | $6.68\times10^{-3}$ | 14.8 |
| 12 | $8_{44} \to 7_{53}$ | 2162.3706 | 0.139 | $1.14\times10^{-2}$ | 15.6 |
| 13 | $9_{37} \to 8_{44}$ | 2531.9175 | 0.118 | $1.87\times10^{-2}$ | 17.9 |

**Table 3b**. Parameters $\chi_{ik}$ and $f_{ik}$ for para-$H_2{}^{16}O$ lines

| No. | Transition $i \to k$ | 400 K | | 1000 K | |
|---|---|---|---|---|---|
| | | $\chi_{ik}$, $10^{-18}$ $cm^2$ | $f_{ik}$, $10^{-2}$ $s^{-1}$ | $\chi_{ik}$, $10^{-18}$ $cm^2$ | $f_{ik}$, $10^{-2}$ $s^{-1}$ |
| | 1 | 2 | 3 | 4 | 5 |
| 1 | $3_{13} \to 2_{20}$ | 12 | 0.4 | 14 | 1.1 |
| 2 | $5_{15} \to 4_{22}$ | 10 | 0.7 | 13 | 2.2 |
| 3 | $6_{42} \to 5_{51}$ | 0.3 | 0.07 | 0.7 | 0.3 |
| 4 | $5_{33} \to 4_{40}$ | 5.4 | 0.7 | 10 | 2.7 |
| 5 | $9_{28} \to 8_{35}$ | 3.7 | 0.4 | 12 | 2.9 |
| 6 | $4_{22} \to 3_{31}$ | - | - | 13 | 1.1 |
| 7 | $5_{24} \to 4_{31}$ | 14 | 0.5 | 35 | 3.0 |
| 8 | $7_{44} \to 6_{51}$ | 0.4 | 0.04 | 3.9 | 0.7 |
| 9 | $7_{26} \to 6_{33}$ | 12 | 0.6 | 29 | 3.1 |
| 10 | $6_{33} \to 5_{42}$ | - | - | 8.5 | 0.8 |
| 11 | $7_{35} \to 6_{42}$ | 3.1 | 0.1 | 17 | 1.5 |
| 12 | $8_{44} \to 7_{53}$ | - | - | 0.3 | 0.02 |
| 13 | $9_{37} \to 8_{44}$ | 0.8 | 0.03 | 7.5 | 0.4 |

Note. To obtain the gain $\gamma_v$ at the line center, the parameter $\chi_{ik}$ should be multiplied by the number density of "working" molecules - para-$H_2{}^{16}O$.

**Table 4a.** Parameters of radiation for ortho-$H_2^{18}O$ lines

| No. | Transition $i \to k$ | Frequency, GHz | $A_{ik}$, s$^{-1}$ | $2\pi S_{ik}/h$, $10^{-10}$ s$^{-1}$ cm$^3$ | $\chi_{ik}$, $10^{-18}$ cm$^2$ | $N_{H_2{}^{16}O}$, $10^{19}$ cm$^{-2}$ | $f_{ik}$, $10^{-2}$ s$^{-1}$ |
|---|---|---|---|---|---|---|---|
| | 1 | 2 | 3 | 4 | 5 | 6 | 7 |
| 1 | $6_{16} \to 5_{23}$ | 5.625105 | $2.95\times10^{-11}$ | 1.76 | 7.0 | 9.5 | 1.0 |
| 2 | $4_{14} \to 3_{21}$ | 390.60799 | $3.14\times10^{-5}$ | 3.86 | 110 | 0.6 | 2.0 |
| 3 | $4_{23} \to 3_{30}$ | 489.05410 | $6.94\times10^{-5}$ | 4.35 | 86 | 0.8 | 2.1 |
| 4 | $5_{32} \to 4_{41}$ | 692.07898 | $1.52\times10^{-4}$ | 4.10 | 4.3 | 15 | 0.2 |
| 5 | $3_{12} \to 2_{21}$ | 1181.3954 | $2.87\times10^{-3}$ | 9.94 | 230 | 0.3 | 0.9 |
| 6 | $6_{34} \to 5_{41}$ | 1216.8483 | $1.62\times10^{-3}$ | 9.49 | 1.3 | 51 | 0.05 |
| 7 | $6_{25} \to 5_{32}$ | 1340.7322 | $2.31\times10^{-3}$ | 10.1 | 23 | 3 | 0.4 |
| 8 | $5_{23} \to 4_{32}$ | 1974.6418 | $1.26\times10^{-2}$ | 14.7 | 7.2 | 9.3 | 0.04 |

**Table 4b**. Parameters of radiation for para-$H_2^{18}O$ lines

| No. | Transition $i \to k$ | Frequency, GHz | $A_{ik}$, s$^{-1}$ | $2\pi S_{ik}/h$, $10^{-10}$ s$^{-1}$ cm$^3$ | $\chi_{ik}$, $10^{-18}$ cm$^2$ | $N_{H_2{}^{16}O}$, $10^{19}$ cm$^{-2}$ | $f_{ik}$, $10^{-2}$ s$^{-1}$ |
|---|---|---|---|---|---|---|---|
| | 1 | 2 | 3 | 4 | 5 | 6 | 7 |
| 1 | $3_{13} \to 2_{20}$ | 203.4074 | $4.79\times10^{-6}$ | 3.25 | 160 | 1.2 | 1.6 |
| 2 | $5_{15} \to 4_{22}$ | 322.46512 | $1.05\times10^{-5}$ | 2.81 | 41 | 4.9 | 0.9 |
| 3 | $5_{33} \to 4_{40}$ | 537.33754 | $6.89\times10^{-5}$ | 3.98 | 1.3 | 160 | 0.08 |
| 4 | $4_{22} \to 3_{31}$ | 970.27469 | $6.76\times10^{-4}$ | 5.42 | 23 | 8.8 | 0.3 |
| 5 | $5_{24} \to 4_{31}$ | 1003.2756 | $9.69\times10^{-4}$ | 8.59 | 39 | 5.1 | 0.5 |
| 6 | $6_{24} \to 5_{33}$ | 3017.1588 | $7.54\times10^{-2}$ | 29.1 | 4.5 | 45 | 0.02 |

Note. For the ortho-$H_2^{18}O$ $6_{16} \to 5_{23}$ transition, the results of our calculations are presented for the most intense hyperfine component. The gas kinetic temperature in the calculations was assumed to be 1000 K. To obtain the gain $\gamma_\nu$ at the line center, the parameter $\chi_{ik}$ should be multiplied by the number density of "working" molecules - ortho-$H_2^{18}O$ or para-$H_2^{18}O$. The parameter $N_{H_2{}^{16}O}$ is the $H_2^{16}O$ column density at which the optical depth in $H_2^{18}O$ lines is equal to unity.

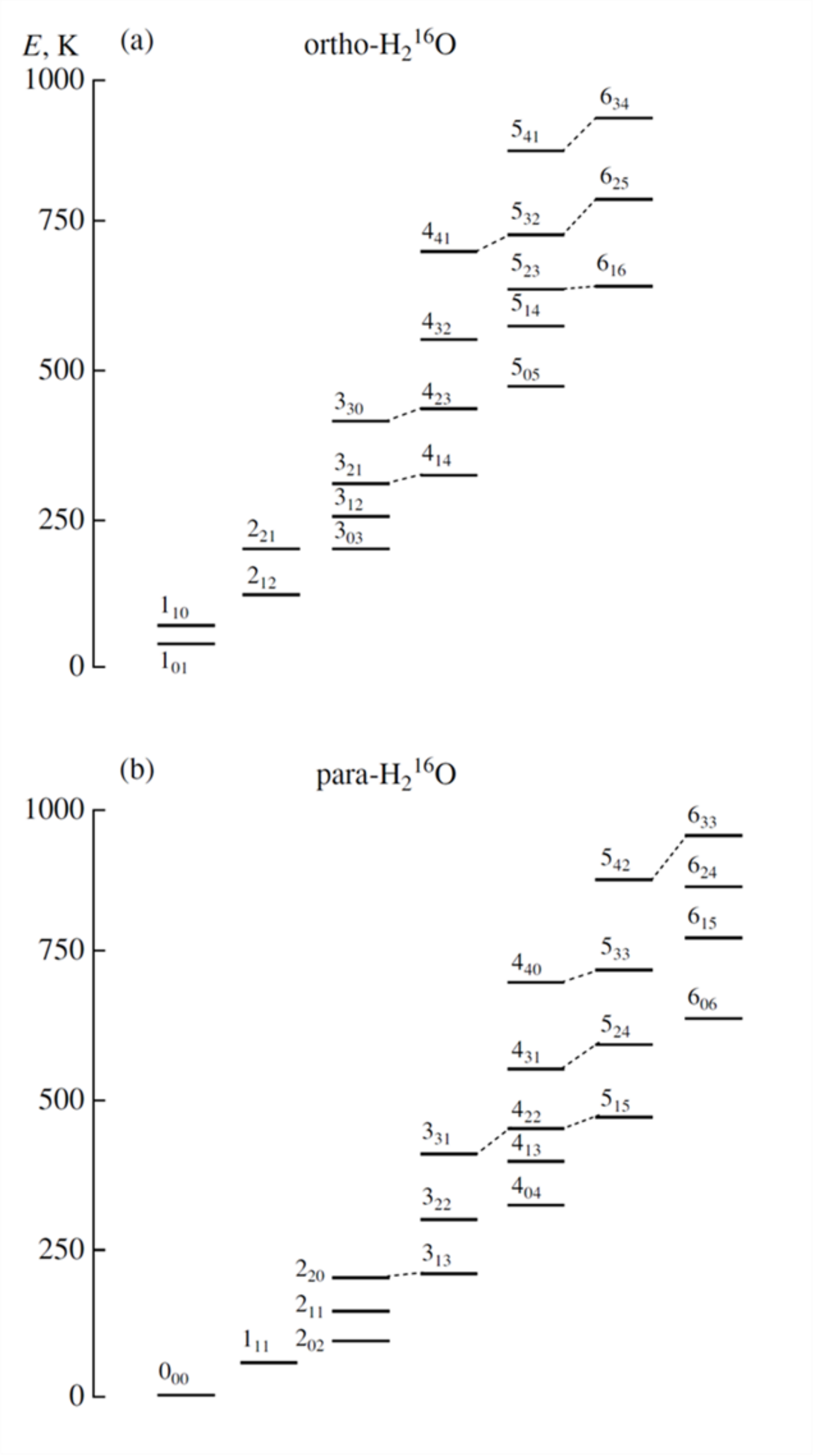

**Fig. 1.** Diagram of lower rotational levels for the ground electronic-vibrational state of ortho-$H_2^{16}O$ (a) and para-$H_2^{16}O$ (b) molecules. The energy scale in kelvins is shown on the left. The dashed lines indicate the transitions for which, according to our calculations, an inverted level population should be observed (Tables 2a and 3a). The diagrams of rotational levels of $H_2^{18}O$ molecules are similar.

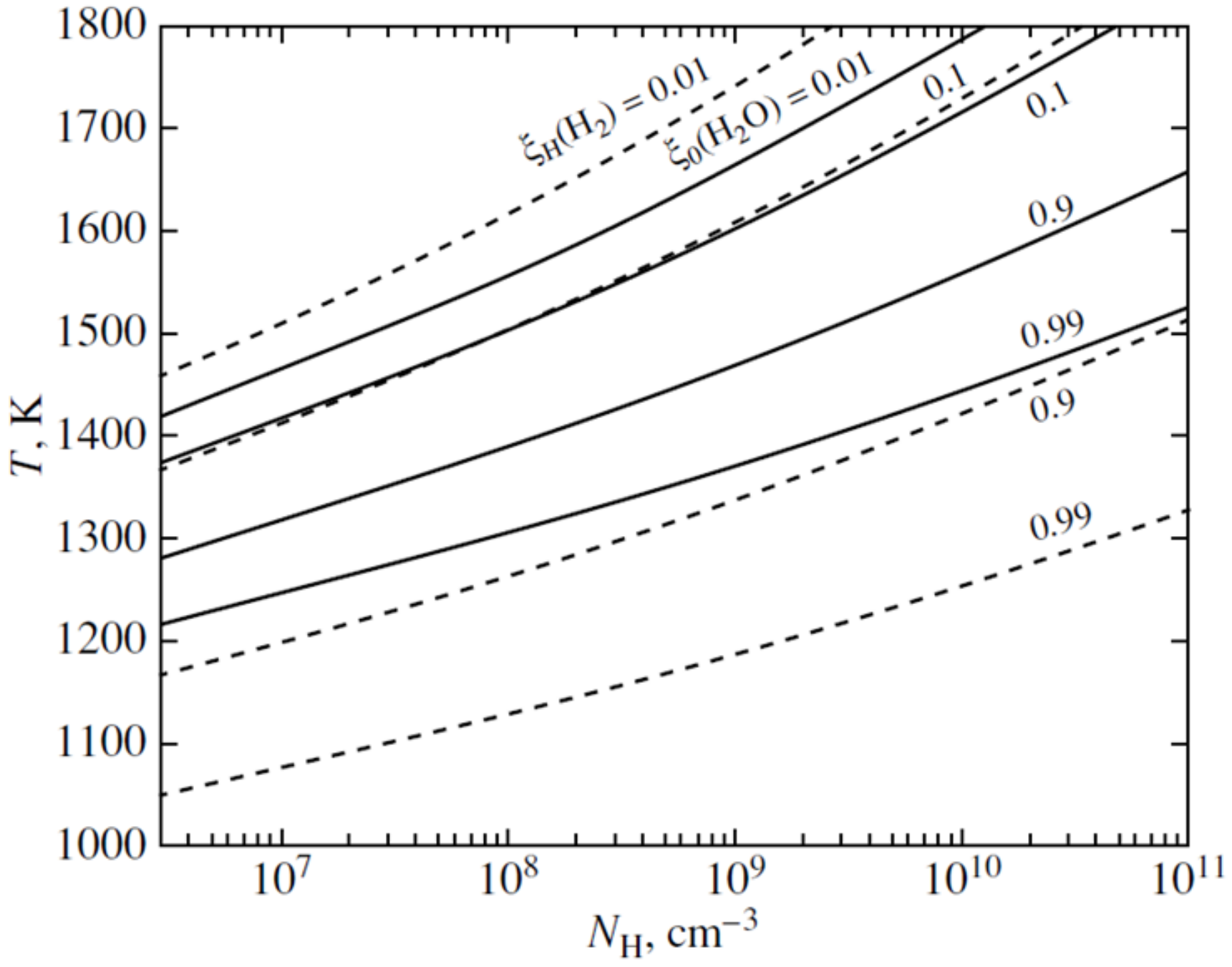

**Fig. 2.** The relative abundances of molecular hydrogen and water molecules as functions of the gas temperature $T$ and the total number density of hydrogen nuclei $N_H$. The solid curves correspond to fixed values of the parameter $\xi_O(H_2O)$, the fraction of oxygen bound in $H_2O$. The dashed curves correspond to fixed values of the parameter $\xi_H(H_2)$, the fraction of hydrogen bound in $H_2$.

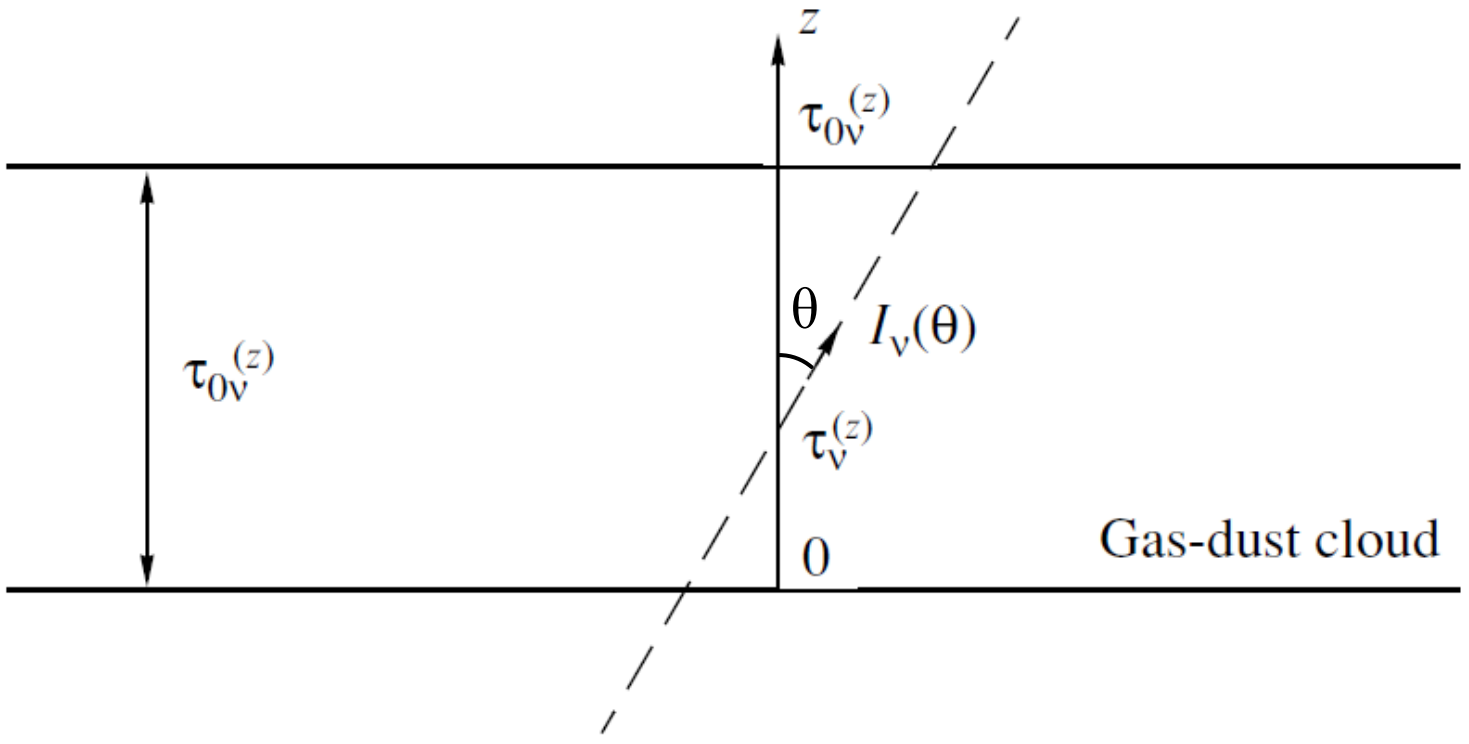

**Fig. 3.** Cloud model. The direction of the $z$ axis coincides with the normal to the cloud plane.

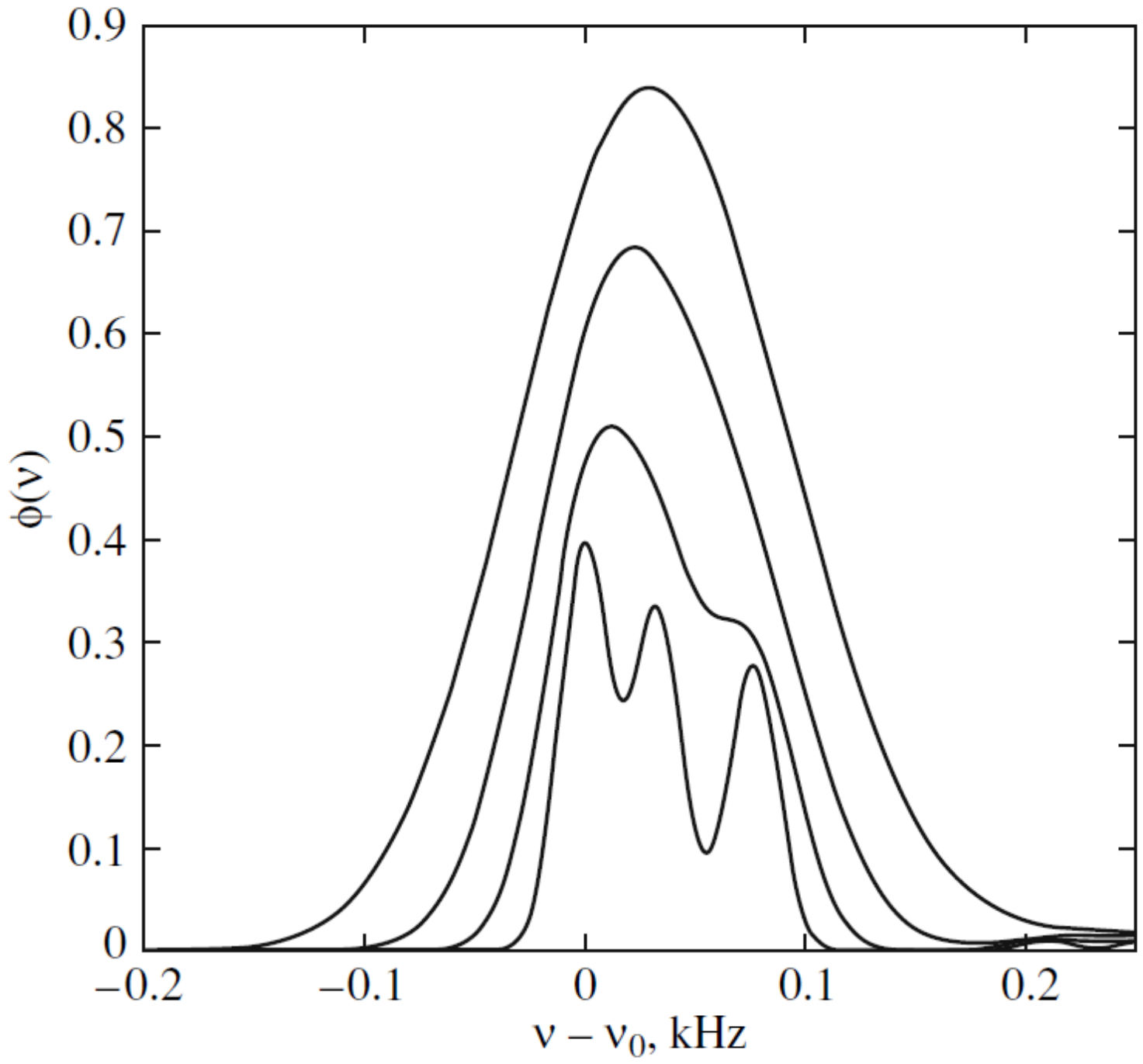

**Fig. 4.** Spectral gain profile in the $6_{16} \rightarrow 5_{23}$ ortho-$H_2{}^{16}O$ line for $T = 50$, 150, 400, and 1000 K (from bottom to top). The frequency shift in kHz relative to the most intense hyperfine component is along the horizontal axis. The profile was normalized in such a way that $\int_{-\infty}^{+\infty} d\nu\, \phi(\nu) = \sqrt{\pi}\, \nu_{ik}\, \upsilon_T / c$; in this case, the profile broadening due to the turbulent motions of gas was disregarded.

*Translated by V. Astakhov*